\documentclass[superscriptaddress,twocolumn,prl,showpacs,preprintnumbers,amsmath,amssymb]{revtex4}

\usepackage{color}
\usepackage{graphicx}
\newcommand{\newatop}[2]{\genfrac{}{}{0pt}{}{#1}{#2}}

\newcommand{\rr}[1]{{\normalfont\textrm{#1}}}
\newcommand{\cc}[1]{{\mathcal{#1}}}
\newcommand{\bb}[1]{{\mathbb{#1}}}
\newcommand{\zero}{{\mathbf{0}}}
\newcommand{\puno}{{\mathbf{+1}}}
\renewcommand{\puno}{{\mathbf{u}}}
\newcommand{\muno}{{\mathbf{-1}}}
\renewcommand{\muno}{{\mathbf{d}}}
\newcommand{\eC}{{\rr{\bf c}^\rr{e}}}
\newcommand{\oC}{{\rr{\bf c}^\rr{o}}}
\newcommand{\cC}{{\rr{\bf c}}}
\newcommand{\cF}{{\rr{\bf f}}}

\begin{document}

%\preprint{Draft: \today}  
\title{Competitive nucleation in reversible Probabilistic Cellular Automata}

\author{Emilio N.M.\ Cirillo}
\affiliation{Dipartimento Me.\ Mo.\ Mat.,  Universit\`a degli Studi di Roma
``La Sapienza," via A.\ Scarpa 16, 00161 Roma, Italy}
%\email{cirillo@dmmm.uniroma1.it}
%\thanks{E.N.M.C.\ thanks Eurandom for the kind hospitality.}

\author{Francesca R.\ Nardi}
\affiliation{Department of Mathematics and Computer Science,
Eindhoven University of Technology,
P.O.\ Box 513, 5600 MB Eindhoven, The Netherlands}
%\email{F.R.Nardi@tue.nl}
\affiliation{Eurandom, P.O.\ Box 513, 5600 MB, Eindhoven, The Netherlands}
%\altaffiliation{Eurandom, P.O.\ Box 513, 5600 MB, Eindhoven, The Netherlands}

\author{Cristian Spitoni}
\affiliation{Eurandom, P.O.\ Box 513, 5600 MB, Eindhoven, The Netherlands}
\affiliation{Mathematical Institute, Leiden University, P.O.\ Box 9512,
2300 RA Leiden, The Netherlands}
%\email{spitoni@eurandom.tue.nl}

%\date{\today}

\begin{abstract}
The problem of competitive nucleation in the framework of Probabilistic 
Cellular Automata is studied from the dynamical point of view. 
The dependence of the metastability scenario on the self--interaction is 
discussed.
An intermediate metastable phase, made of two 
flip--flopping chessboard configurations, shows up depending 
on the ratio between the magnetic field and the self--interaction.
A behavior similar to the one of the stochastic Blume--Capel model 
with Glauber dynamics is found. 
\end{abstract}



\maketitle

%%%%%%%% Il lavoro
%\newpage
%\section{Introduction}
%\label{s:intro}
%\par\noindent
%\section{The model}
%\label{s:modello}
%\par\noindent
Metastable states are common in nature; they show up 
in connection with first order phase transitions. Well known examples are 
super--cooled and super--heated liquids. Their statistical mechanics 
description revealed to be a challenging task. 
An approach based on equilibrium states has been 
developed via analytic continuation techniques \cite{[L]}
and via the introduction of equilibrium systems on suitably restricted 
sets of configurations \cite{[PL1],[BLLM],[LL]}.
The purely dynamical point of view, dating back to
Ref.\ \cite{[CGOV]}, has been developed via the pathwise technique 
\cite{[OV]} and the potential theoretical approach \cite{[BEGK]}.

We shall stick to the dynamical description 
to investigate competing metastable states. 
This situation arise in many physical processes,
such as the crystalization of proteins \cite{[tWF],[LN]}
and their approach to equilibrium \cite{[TW]}. 
The extreme situation 
is represented by the glasses, in which the presence of a huge number 
of minima of the energy landscape prevents the system from reaching the 
equilibrium \cite{[BK]}.
The study of these systems is difficult, since 
the minima of the energy and the decay pathways between them 
change when the control parameters are varied. 
It is then of interest the study of models in which  
a complete control of the variations induced on the energy landscape by 
changes in the parameters is possible.

In this perspective,
the analysis of the Blume--Capel model in Ref.\ \cite{[CO],[FGRN]} and 
that of the Potts model in Ref.\ \cite{[SLL]} are of great interest.
In the Blume--Capel model the sites of the lattice can be either empty or 
occupied by a $1/2$--spin particle. 
The interaction favors the presence of neighboring aligned spins; 
the chemical potential $\lambda$ controls the 
tendency to have particles or lacunas on the lattice and 
the magnetic field $h$, depending on its sign, favors either the 
pluses or the minuses.
Depending on the parameters, in the zero temperature limit
the stable state is the one with all the spins up ($\puno$) 
or all the spins down ($\muno$) or no particle at all ($\zero$). 
Let $h,\lambda>0$, so that the unique stable state 
is $\puno$, and set $a=h/\lambda$. 
For $a<1$ the transition from the metastable state 
$\muno$ to $\puno$ is achieved via a sequence of increasing plus square 
droplets in the sea of minuses. 
For $1<a<2$ and $h$ small, the transition from $\muno$ to $\puno$ is realized
via increasing squared frames in which 
the internal pluses are separated by the external minuses by a frame
of zeros large one. 
For $a>2$ and $h$ small, the system started at $\muno$ visits the  
state $\zero$ before reaching $\puno$; 
the transition from $\muno$ to $\zero$ is achieved 
via increasing zero square droplets in the sea of minuses,
while the transition from $\zero$ to $\puno$ is realized
via increasing plus square droplets in the sea of zeros. 

%In the Potts model, the spin associated with each site of the lattice can
%assume $n$ values. 
%The interaction favors the presence of neighboring spins of the same species;
%moreover, an external field selects the preferred state. By choosing properly
%the interaction among the different species, it is possible to vary 
%the energy landscape in order to have a prescribed sequence of 
%metastable states. In Ref.\ \cite{[SLL]} some of the possible situations
%are studied numerically. 

We study, here, metastability for a Probabilistic 
Cellular Automaton \cite{[D]} with self--interaction $\kappa$, focusing 
on the dependence of the metastability scenario on such a parameter.
The model interpolates those 
studied in Ref.\ \cite{[CN]} ($\kappa=0$) and \cite{[CNS],[BCLS]} ($\kappa=1$).
For $\kappa=0$ each spin 
interacts only with its nearest neighbors; for $\kappa=1$
the self--interaction has the same strength as the nearest neighbor coupling. 
In absence of self--interaction an intermediate 
metastable state shows up; it is proven that the intermediate 
state is visited during the transition from the 
metastable to the stable state. 
The role played by the intermediate state changes 
as the self--interaction $\kappa$ is varied. 
Quite surprisingly, results 
similar to those found in Ref.\ \cite{[CO]} for the 
Blume--Capel model are obtained.

Consider the two--dimensional torus $\Lambda=\{0,\dots,L-1\}^2$,
with $L$ even, endowed with the Euclidean metric;
$x,y\in\Lambda$ are \textit{nearest neighbors} iff
their mutual distance is equal to $1$.
Associate a variable $\sigma(x)=\pm1$
with each site $x\in\Lambda$ and let $\cc{S}=\{-1,+1\}^{\Lambda}$ be the 
\textit{configuration space}.
Let $\beta>0$ and $\kappa,h\in[0,1]$.
Consider the Markov chain $\sigma_n$, with $n=0,1,\dots$,
on $\cc{S}$ with {\it transition matrix}
\begin{equation}
\label{markov}
p(\sigma,\eta)
=\prod_{x\in\Lambda}p_{x,\sigma}\left(\eta(x)\right)\;\;\;
\forall\sigma,\eta\in\cc{S}
\end{equation}
where, for $x\in\Lambda$ and $\sigma\in\cc{S}$,
$p_{x,\sigma}(\cdot)$ is the probability measure on $\{-1,+1\}$
defined as 
%\begin{equation}
%\label{rule}
$p_{x,\sigma}(s)
 =1/[1+\exp\left\{-2\beta s(S_\sigma(x)+h)\right\}]$
%=
%\frac{1}{2}
%\left[1+s\tanh\beta \left(
%S_\sigma(x)+h\right)\right]
%\end{equation}
with $s\in\{-1,+1\}$ and
$S_{\sigma}(x)=\sum_{y\in\Lambda}K(x-y)\,\sigma(y)$
where 
$K(x-y)$ is $0$ if $|x-y|\ge2$,
$1$ if $|x-y|=1$, and
$\kappa$ if $|x-y|=0$.
%The normalization condition 
%$p_{x,\sigma}(s)+p_{x,\sigma}(-s)=1$ is trivially satisfied.
The probability $p_{x,\sigma}(s)$ for the spin $\sigma(x)$ to be equal to $s$
depends only on the values of the spins of $\sigma$ 
in the five site cross centered at $x$.
The metastable behavior of model (\ref{markov}) has been studied
in Ref.\ \cite{[CN]} for $\kappa=0$ and in Ref.\ \cite{[BCLS],[CNS]} for
$\kappa=1$.  

The Markov chain (\ref{markov}) is a
\textit{probabilistic cellular automata};
the chain $\sigma_n$, with $n=0,1,\dots$,
updates all the spins simultaneously and independently at any time.
The chain is \textit{reversible}, see Ref.\ \cite{[D]}, 
with respect to the Gibbs measure
$\mu(\sigma)=\exp\{-\beta H(\sigma)\}/Z$
with
$Z=\sum_{\eta\in\cc{S}}\exp\{-\beta H(\eta)\}$
and
\begin{equation}
\label{ham}
H(\sigma)=
%H_{\beta,h}(\sigma)=
-h\sum_{x\in\Lambda}\sigma(x)
-\frac{1}{\beta}\sum_{x\in\Lambda}\log\cosh\left[\beta
\left(
S_{\sigma}(x)+h\right)\right]
\end{equation}
that is \textit{detailed balance}
$p(\sigma,\eta)\,e^{-\beta H(\sigma)}=
 p(\eta,\sigma)\,e^{-\beta H(\eta)}$
holds for $\sigma,\eta\in\cc{S}$;
hence, $\mu$ is stationary.
We  refer to
$1/\beta$ as to the {\it temperature} and to $h$ as to the
{\it magnetic field};  
the interaction is short range and it is possible to extract the
potentials as described in Ref.\ \cite{[BCLS]}.

Although the dynamics is reversible w.r.t.\ the Gibbs measure associated
to the Hamiltonian (\ref{ham}), the probability 
$p(\sigma,\eta)$ cannot be expressed in terms of  
$H(\sigma)-H(\eta)$, as usually happens for Glauber dynamics. 
Given $\sigma,\eta\in\cc{S}$, we define the \textit{energy cost} 
\begin{equation}
\label{defdelta}
\Delta(\sigma,\eta)=
 -\lim_{\beta\to\infty}\frac{\log p(\sigma,\eta)}{\beta}
 =
\!\!\!\!\!\!\!\!\!
\sum_{\newatop{x\in\Lambda:}{\eta(x)[S_\sigma(x)+h]<0}}
\!\!\!\!\!\!\!\!\!
2|S_\sigma(x)+h|
\end{equation}
Note that $\Delta(\sigma,\eta)\ge0$ and $\Delta(\sigma,\eta)$ is not
necessarily equal to $\Delta(\eta,\sigma)$;
it can be proven, see \cite[Section~2.6]{[CNS]}, that 
\begin{equation}
\label{cri01}
e^{-\beta\Delta(\sigma,\eta)-\beta\gamma(\beta)}
\le
p(\sigma,\eta)
\le
e^{-\beta\Delta(\sigma,\eta)+\beta\gamma(\beta)}
\end{equation}
with $\gamma(\beta)\to0$ in the zero temperature limit $\beta\to\infty$.
Hence, $\Delta$ can be 
interpreted as the cost of the transition from $\sigma$ to $\eta$
and plays the role that, in the context of Glauber dynamics, is 
played by the difference of energy.

To pose the problem of metastability 
it is necessary to understand the structure of the ground states; 
since the Hamiltonian depends on $\beta$, their definition deserves 
some thinking. The ground states are those configurations on which the Gibbs
measure $\mu$ concentrates when
$\beta\to\infty$; hence, they can be defined as the
minima of the \textit{energy}
\begin{equation}
\label{hl}
E(\sigma)=
\lim_{\beta\to\infty}H(\sigma)
=
-h\sum_{x\in\Lambda}\sigma(x)
-\sum_{x\in\Lambda}|S_{\sigma}(x)+h|
\end{equation}
For $\cc{X}\subset\cc{S}$, we set 
$E(\cc{X})=\min_{\sigma\in\cc{X}}E(\sigma)$.
For $h>0$ the configuration $\puno$, with
$\puno(x)=+1$ for $x\in\Lambda$, is the unique ground state,
indeed each site contributes to the energy with $-h-(4+\kappa+h)$. 
For $h=0$, the ground states are the configurations such that all the sites
contribute to the sum (\ref{hl}) with $4+\kappa$. 
Hence, for $\kappa\in(0,1]$, the sole ground states are the configurations 
$\puno$ and $\muno$, with $\muno(x)=-1$ for $x\in\Lambda$.
For $\kappa=0$, the configurations $\eC,\oC\in\cc{S}$ such that 
$\eC(x)=(-1)^{x_1+x_2}$ and $\oC(x)=(-1)^{x_1+x_2+1}$
for $x=(x_1,x_2)\in\Lambda$ are ground states, as well. Notice that 
$\eC$ and $\oC$ are chessboard--like states with the pluses  
on the even and odd sub--lattices, respectively; we set $\cC=\{\eC,\oC\}$. 
Since the side length $L$ of the torus $\Lambda$ 
is even, then $E(\eC)=E(\oC)=E(\cC)$.

We study those energies as a function of $\kappa$ and $h$, recalling
that periodic boundary conditions are considered. 
We have 
$E(\puno)=-L^2(4+\kappa+2h)$, 
$E(\muno)=-L^2(4+\kappa-2h)$, 
and
$E(\cC)=-L^2(4-\kappa)$;
hence
$E(\cC)>E(\muno)>E(\puno)$ for $0<h<\kappa\le1$, 
$E(\cC)=E(\muno)>E(\puno)$ for $0<h=\kappa\le1$, 
and
$E(\muno)>E(\cC)>E(\puno)$ for $0<\kappa<h\le1$.

We can now pose the problem of metastability 
at finite volume and temperature tending to zero
(Friedlin--Wentzel regime).
Following Ref.\ \cite{[OV]}, see also Ref.\ \cite[Appendix]{[CNS]}, 
given a sequence of configurations
$\omega=\omega_1,\dots,\omega_n$, with $n\ge2$, we define the 
\textit{energy height} along the path $\omega$ as  
$\Phi_\omega=\max_{i=1,\dots,|\omega|-1}
        [E(\omega_i)+\Delta(\omega_i,\omega_{i+1})]$.
Note that the definition does not depend on the direction in which 
the path $\omega$ is followed. More precisely, denoted by $\omega'$ 
the path $\omega_n,\omega_{n-1},\dots,\omega_1$, 
since 
\begin{equation}
\label{eee}
E(\sigma)+\Delta(\sigma,\eta)=E(\eta)+\Delta(\eta,\sigma)
\end{equation}
for any $\sigma,\eta\in\cc{S}$, it follows that 
$\Phi_\omega=\Phi_{\omega'}$; (\ref{eee}) is
consequence of the detailed balance principle.
Given $A,A'\subset\cc{S}$,
%we denote by $\Theta(A,A')$ the set of all the paths starting in $A$ and 
%ending in $A'$ and 
we let the \textit{communication energy} between $A$ and $A'$ be 
the minimal energy height $\Phi_\omega$ over the set of paths $\omega$
starting in $A$ and ending in $A'$.
%$\Phi(A,A')=\min_{\omega\in\Theta(A,A')} \Phi_\omega$.
For any $\sigma\in\cc{S}$, we let
%$\cc{I}_\sigma=\{\eta\in\cc{S}:\,E(\eta)<E(\sigma)\}$ be the set of states 
$\cc{I}_\sigma\subset\cc{S}$
be the set of configurations with energy strictly below $E(\sigma)$ and
$V_\sigma=\Phi(\sigma,\cc{I}_\sigma)-E(\sigma)$ be the \textit{stability level
of} $\sigma$, that is the energy barrier that, starting from $\sigma$, must be 
overcome to reach the set of configurations with energy smaller than 
$E(\sigma)$; we set $V_\sigma=\infty$ if $\cc{I}_\sigma=\emptyset$.
We denote by $\cc{S}^\rr{s}$ the set of
global minima of the energy (\ref{hl}), namely, the collection of the 
ground states, and suppose that 
the \textit{communication energy}
$\Gamma=\max_{\sigma\in\cc{S}\setminus\cc{S}^\rr{s}}V_\sigma$
is strictly positive.
Finally, we define the set of \textit{metastable states}
$\cc{S}^\rr{m}=\{\eta\in\cc{S}:\,V_\eta=\Gamma\}$.
The set $\cc{S}^\rr{m}$ deserves its name, 
since it is proven the following (see, e.g., Ref.\ \cite[Theorem~A.2]{[CNS]}):
pick $\sigma\in\cc{S}^\rr{m}$, consider the 
chain $\sigma_n$ started at 
$\sigma_0=\sigma$, then the \textit{first hitting time}
$\tau_{\cc{S}^\rr{s}}=\inf\{t>0:\,\sigma_t\in\cc{S}^\rr{s}\}$ to the 
ground states is a random variable with mean exponentially large in $\beta$,
that is 
\begin{equation}
\label{mnos4.9}
\lim_{\beta\to\infty}
\frac{1}{\beta}\,\log\bb{E}_\sigma[\tau_{\cc{S}^\rr{s}}]=\Gamma
\end{equation}
with $\bb{E}_\sigma$ the average on the trajectories 
started at $\sigma$.

In this regime the description of metastability is reduced
to the computation of $\cc{S}^\rr{s}$, $\Gamma$, and $\cc{S}^\rr{m}$.
We choose the parameters of the model (\ref{markov}) in such a way that 
$0<h<1$, $h\neq\kappa$, and $2/h$, $2/(h-\kappa)$, $2/(h+\kappa)$,
and $(2+\kappa-h)/h$ are not integer.
The configuration $\puno$ 
is then the unique ground state, i.e., $\cc{S}^\rr{s}=\{\puno\}$.
Two candidates for metastability are $\muno$ and $\cC$;
to find $\cc{S}^\rr{m}$, one should 
compute $\Gamma$ and prove that either $V_\muno$ or 
$V_\cC$ is equal to $\Gamma$. 
This is a difficult task, indeed all the paths $\omega$ 
connecting $\muno$ and $\cC$ to $\puno$ must be taken into account
and the related energy heights $\Phi_\omega$ computed.
Since at each time step all the spins of the lattice can be updated, 
the structure of the trajectories is highly complicated.
This is why the study of the energy landscape 
of probabilistic cellular automata is very difficult
\cite[Theorem~2.3]{[CNS]}; such a task is 
simpler for serial Glauber dynamics,
where a sort of general approach can be developed
\cite[Section~7.6]{[OV]}.

We develop an heuristic argument to compute $\Gamma$.
Recall (\ref{defdelta}) 
%that the energy cost of a transition from $\sigma$ to $\eta$ is 
%represented by $\Delta(\sigma,\eta)$ for any $\sigma,\eta\in\cc{S}$. 
and note that $\kappa$ and $h$ have been chosen so that 
$S_\sigma(x)+h\neq0$.
Thus, it follows that, given $\sigma\in\cc{S}$, 
there exists a unique $\eta\in\cc{S}$ such that $\Delta(\sigma,\eta)=0$;
the configuration $\eta$ is such that  
$\eta(x)[S_\sigma(x)+h]>0$ for all $x\in\Lambda$
and is the unique configuration to which the system can jump, starting 
from $\sigma$, with probability tending to one in the limit
$\beta\to\infty$ (see (\ref{cri01})). 
We say that $\sigma\in\cc{S}$ is a \textit{local minimum} 
of the energy iff $\Delta(\sigma,\sigma)=0$; 
starting from a local minimum, transitions to  
different configurations have strictly positive energy cost
and thus happen with negligible probability in the zero temperature limit.
It is immediate that $\muno$ and $\puno$ are local minima of the energy,
while $\eC$ and $\oC$ are not, indeed 
$\eC(x)[S_\eC(x)+h]<0$ and $\oC(x)[S_\oC(x)+h]<0$ for all $x\in\Lambda$.
We also have that $\Delta(\eC,\oC)=\Delta(\oC,\eC)=0$, hence 
at very low temperature, the system started in $\oC$ is trapped 
in a continuous flip--flop between $\oC$ and $\eC$.
A peculiarity of parallel dynamics is the existence 
of pairs $\sigma,\eta\in\cc{S}$ in which the chain is trapped since 
$\Delta(\sigma,\eta)=\Delta(\eta,\sigma)=0$;
the probability to exit such a pair is exponentially small in $\beta$. 

We characterize, now, the local minima and the trapping pairs.
For what concerns the local minima, we consider a 
configuration $\sigma$ and study the sign of $S_\sigma(x)+h$.
Suppose, first, $h<\kappa$ and recall $\kappa\le1$; the sign of 
$S_\sigma(x)+h$ equals the sign of the majority of the spins in the 
five site cross centered at $x$.
Hence, $\sigma$ is a local minimum iff for each site $x$ there exist at
least two nearest neighbors such that the associated spins are equal to 
$\sigma(x)$.
Suppose, now, $h>\kappa\ge0$; the sign of $S_\sigma(x)+h$ is negative
iff at least three among the spins associated to neighboring sites of $x$ 
are minus.
Hence, $\sigma$ is a local minimum iff for each site $x$ such that 
$\sigma(x)=-1$ there exist at least three negative minus neighbors and 
for each site $x$ such that $\sigma(x)=+1$ there exist at least two positive 
neighbors.
In conclusion, for $h>\kappa$ the local minima of the energy are those 
configurations in which all the pluses, if any, are precisely those 
associated with the sites inside a rectangle (\textit{plus--minus} droplets). 
For $h<\kappa$ the local minima 
are all the configurations that can be drawn adding 
pluses to $\muno$ so that each plus (resp.\ minus) has at least 
(resp.\ at most) two neighboring pluses.
Plus--minus rectangular droplets are local minima also in this case.
For what concerns the trapping pairs,
consider a configuration $\sigma$ with a rectangle of chessboard
plunged in the sea of minuses (\textit{chessboard--minus} droplet)
and let $\eta$ be the configuration 
obtained flipping all the spins associated with sites in the chessboard 
rectangle.
The configuration $\sigma,\eta$ form a trapping pair only for $h>\kappa$.
Indeed, it is immediate to show that all the spins of the chessboard 
tend to flip, some thinking is necessary only for the minus corners.
Let $x$ be the corner site with $\sigma(x)=-1$, 
since $S_\sigma(x)+h=-\kappa+h$, we have that $S_\sigma(x)+h>0$ for $h>\kappa$ 
and $S_\sigma(x)+h<0$ for $h<\kappa$. Thus, the spin tends to flip in the 
former case and not in the latter. 

The local minima and the trapping pairs can be used to construct 
the optimal paths connecting $\muno$ and $\cC$ to the ground state $\puno$.
We distinguish two cases.

\textit{Case $h>\kappa\ge0$.\/} 
Although $\eC$ and $\oC$ are not local minima of the energy, 
the system started in $\cC$ is trapped in a continuous flip--flop 
between $\oC$ and $\eC$. This trapping persists even if a
rectangle of pluses is inserted in the chessboard background 
(\textit{plus--chessboard} droplet);
a path from $\cC$ to $\puno$ can be constructed with a sequence of 
such droplets.
The difference of energy between two plus--chessboard droplets
with side lengths respectively given by $\ell,m\ge2$ and  $\ell,m+1$ is  
equal to $4-2(\kappa+h)\ell$.
It then follows that the energy of a such a droplet is increased by 
adding an $\ell$--long slice iff 
$\ell\ge\lfloor2/(\kappa+h)\rfloor+1=\lambda^\puno_\cC$
($\lfloor x\rfloor$ denotes the largest integer smaller than the 
real $x$).
The length $\lambda^\puno_\cC$ is called the 
\textit{critical length}.
It is reasonable that the energy barrier $V_\cC$ is given
by the difference of energy between 
the smallest supercritical plus--chessboard droplet,
i.e., the plus--chessboard square droplet with side length 
$\lambda^\puno_\cC$, and the configuration $\cC$;
by using (\ref{hl}) we get that
such a difference of energy 
is equal~\cite{energia} 
to $\Gamma^\puno_\cC=8/(\kappa+h)$.

A path from $\muno$ to $\puno$ can be constructed with a sequence of 
plus--minus droplets.
By using (\ref{hl}) we get that the difference of energy 
between two plus--minus droplets with side lengths respectively given 
by $\ell,m\ge2$ and  $\ell,m+1$ is $4(2-h\ell)$.
It then follows that the energy of a plus--minus droplet is increased by 
adding an $\ell$--long slice iff 
$\ell\ge\lfloor2/h\rfloor+1=\lambda^\puno_\muno$.
The length $\lambda^\puno_\muno$ is the 
critical length for the plus--minus droplets;
by using (\ref{hl}) we get that the difference of 
energy between the smallest supercritical plus--minus droplet 
and $\muno$ is equal to $\Gamma^\puno_\muno=16/h$.

An alternative path from $\muno$ to $\puno$ can be constructed via a 
sequence of \textit{frames} with the internal rectangle of pluses 
separated by the external minuses by a stripe 
of chessboard large one. 
These are peculiar trapping pairs in which the flip--flopping spins
are those associated with the sites in the stripe of chessboard.
We can prove that the difference of energy 
between two frames with internal (rectangle of pluses) side lengths 
respectively given by $\ell,m\ge2$ and  $\ell,m+1$ is equal to 
$8-4(h-\kappa)-4h\ell$, so that the critical length for those frames 
is given by 
$\lambda^\cF_\muno=\lfloor(2-h+\kappa)/h\rfloor+1$ and 
the difference of energy between the smallest supercritical frame 
and $\muno$ is equal 
to $\Gamma^\cF_\muno=16[1-(h-\kappa)/2]^2/h$.

A path from $\muno$ to $\cC$ can be constructed with a sequence of 
chessboard--minus droplets.
By using (\ref{hl}) we get that the difference of energy 
between two chessboard--minus droplets with side lengths respectively given 
by $\ell,m\ge2$ and  $\ell,m+1$ is  equal to $4-2(h-\kappa)\ell$.
It then follows that the energy of a chessboard--minus droplet is increased by 
adding an $\ell$--long slice iff 
$\ell\ge\lfloor2/(h-\kappa)\rfloor+1=\lambda^\cC_\muno$.
The length $\lambda^\cC_\muno$ is the 
critical length for the chessboard--minus droplets;
the energy difference of energy between the smallest supercritical 
chessboard--minus droplet and $\muno$ is equal to 
$\Gamma^\cC_\muno=8/(h-\kappa)$.

Note that $\Gamma^\cF_\muno<\Gamma^\puno_\muno$ for $h,\kappa$ small.
Moreover, let $a=h/\kappa$ and remark that, provided the magnetic field 
$h$ is chosen small enough as a function of $a$,
$\Gamma^\cC_\muno<\Gamma^\cF_\muno$ for $a>2$ and 
$\Gamma^\cC_\muno>\Gamma^\cF_\muno$ for $1<a<2$.
Hence, for $a>2$ we get $V_\muno=\Gamma^\cC_\muno$, that is the 
chain escapes from $\muno$ and reaches the state $\cC$ 
in a time that can be estimated as in (\ref{mnos4.9})
with $\Gamma=\Gamma^\cC_\muno$. Starting 
from $\cC$ the chain will reach $\puno$ by overcoming the energy 
barrier $V_\cC=\Gamma^\puno_\cC<V_\muno$. Note that $V_\cC=V_\muno$ 
in the limiting case $\kappa=0$, hence both $\cC$ and $\muno$ are metastable 
states (results in \cite{[CN]} are recovered).
For $1<a<2$, $V_\muno=\Gamma^\cF_\muno$, that is the 
chain escapes from $\muno$ and reaches the state $\puno$ via a 
sequence of increasing frames
in a time estimated as in (\ref{mnos4.9}) with 
$\Gamma=\Gamma^\cF_\muno$.

\textit{Case $h<\kappa\le1$.\/} By paying the smallest energy cost any 
local minimum can be transformed in a configuration with the pluses
forming well separated rectangles (see \cite{[BCLS]}); hence,
the most relevant local minima are the plus rectangular droplets. 
As noted above, for this choice of the parameters the system cannot be trapped 
in chessboard--minus droplets.
Thus, the energy barrier $V_\muno$ is given
by the energy $\Gamma^\puno_\muno$ of the smallest supercritical plus droplet. 
As before, we also have $V_\cC=\Gamma^\puno_\cC$.
Since $V_\cC<V_\muno$, we have that $\muno$ is the unique metastable state, the 
communication energy is $\Gamma=\Gamma^\puno_\muno$, the tunneling time 
is $\exp\{\beta \Gamma^\puno_\muno\}$ in the sense (\ref{mnos4.9}), and the 
zero temperature limit transition from the metastable state $\muno$ 
to the stable state $\puno$ is achieved via the nucleation of a plus--minus 
square droplet with side length $\lambda^\puno_\muno$.
For $\kappa=1$ the results proven in \cite{[CNS]} are recovered.

\begin{figure}[t]
\vskip 0.5 cm
\begin{center}
\includegraphics[height=6cm,angle=0]{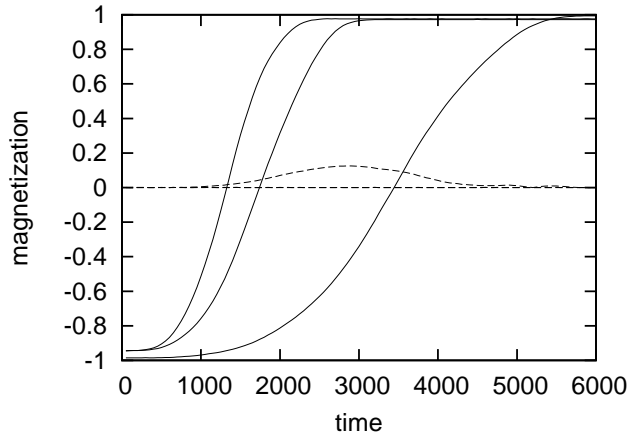}
\caption{The time unit is the time step of the chain.
Solid lines (from the left to the right) represent the magnetization 
of the runs $(\kappa,\beta)=(0.15,0.55),(0.4,0.5),(0,025,0.7)$.
Dashed lines represent the absolute value of the staggered magnetization; 
the non--null curve is found for $(\kappa,\beta)=(0.025,0.7)$.}
\label{f:mag}
\end{center}
\vskip -1 cm
\end{figure}

The metastability scenario depends 
on the ratio between the magnetic field and the self--interaction. 
For $\kappa=0$ the two states $\muno$ and $\cC$ are both metastable.
For $a>2$ and $h$ small, $\cC$ is crucial, although not metastable, 
since it is visited during the transition from the metastable 
state $\muno$ to the stable state $\puno$. 
For $2>a>1$ and $h$ small, the chessboard configuration plays no role at all
and the exit from the metastable $\muno$ state is achieved via the direct 
formation of the plus phase via a sequence of increasing frames.
For $1>a$, the exit from the metastable $\muno$ state is achieved via the 
direct formation of the plus phase via a sequence of increasing  
plus--minus droplets.
The scenario is very similar to the one 
proven in Ref.\ \cite{[CO]} for the Blume--Capel model 
with Glauber (serial) dynamics; the role of the chemical potential $\lambda$ 
is played here by the self--interaction $\kappa$.
This behavior has been tested at finite temperature via a Monte Carlo 
simulation
\footnote{Simulations performed on the Sun Fire X2100 M2 Cluster 
of the Dipartimento Me.\ Mo.\ Mat., 
Universit\`a degli Studi di Roma ``La Sapienza."}.
We have considered $L=1000$, $h=0.2$, and run the chain for  
$(\kappa,\beta)=(0.025,0.7)$, $(0.15,0.55)$, $(0,4,0.5)$.
By measuring the staggered and the usual magnetization, we point out
that the system visits $\cC$ before reaching $\puno$
only in the run $\kappa=0.025$ and $\beta=0.7$ (see Figure~\ref{f:mag}),
which is the only run with $a>2$.


\begin{thebibliography}{00000}
\bibitem{energia}
From now on energy--like quantities are computed neglecting terms being
$O(1)$ for $h,\kappa$ small.
\end{thebibliography}
\end{document}